# Einstein on the Impossibility of Superluminal Velocities

Galina Weinstein

In 1907 Einstein discussed with Wilhelm Wien the occurrences of superluminal velocities in dispersive and absorptive media. Einstein tried to present to Wien an expression for the group velocity in dispersive media that was based on his 1905 relativistic addition theorem for velocities, which he claimed to be valid for absorptive media, and to demonstrate the impossibility of superluminal velocities. However, he recognized that his expression needed an amendment and correction. Einstein did not convince Wien, and was finally confused, because he did not have the correct expression for the group velocity in dispersive media. However, he wrote Wien, "it is beyond doubt that our electromagnetic theories of dispersion can never yield superluminal velocity for the propagation of an optical signal".

In his 1905 relativity paper, "On the Electrodynamics of Moving Bodies", Einstein obtained the addition theorem for velocities.[1]

Along the $X$ axis of the system $K$, a point is moving in the system $k$ with a velocity $v$ according to the equations:

$\xi = w_\xi \tau$, $\eta = w_\eta \tau$, $\zeta = 0$, where, $w_\xi$ and $w_\eta$ denote constants.

Einstein required the motion of the point relative to the system $K$. With the help of the Lorentz transformations, Einstein obtained for the equations of motion of the point:

$$x = \frac{t(w_\xi + v)}{(1 + vw_\xi/c^2)}, y = \frac{w_\eta t\sqrt{1 - v^2/c^2}}{(1 + vw_\xi/c^2)}, z = 0$$

and:

$U^2 = (dx/dt)^2 + (dy/dt)^2$, $w^2 = w^2_\xi + w^2_\eta$

Einstein finally obtained U if $w$ has the direction of the $X$-axis:

$U = (v + w)/(1 + vw/c^2)$.[2]

Einstein concluded, "It follows from this equation that from a composition of two velocities which are smaller than $c$, there always results a velocity smaller than $c$. [U is smaller than c…] It also follows that the velocity of light $c$ cannot be changed by

---

composition with a velocity less than that of light" [U = c always for anything moving with speed c]".[3]

It soon turned out that Einstein's argument concerning the addition theorem for velocities led to the discussion of superluminal velocities. From summer 1907, in six letters to Professor Wilhelm Wien, Einstein discussed the occurrences of velocities exceeding the speed of light *in dispersive and absorptive media*. Einstein tried to answer the question whether such *phase velocities* in dispersive and absorptive media are the physically meaningful *signal velocities* relativity theory requires to be less than or equal to the velocity of light. Though failing to convince Wien, Einstein letters refine and expand upon his published treatments of the possibility of superluminal speeds.[4]

Three years earlier Wien defended the impossibility of superluminal speeds in a debate with Max Abraham. Abraham obtained the expression for the variation of mass with velocity, while assuming rigid spherical electrons keeping their spherical form at any velocity. His spherical electron seemed to permit superluminal velocities, because it was a rigid spherical electron and could be accelerated indefinitely. Hendrik Antoon Lorentz assumed that the moving electron was contracted in the direction of motion and according to Lorentz's theory it would require an infinite amount of energy to accelerate his deformable electron to the speed of light. Therefore Lorentz's model of the electron did not allow superluminal speeds at all. Wien was thus a supporter of Lorentz's electron. Einstein's 1905 relativity theory also did not allow superluminal speeds at all.[5]

By this time in his June 1907 review article on the theory of relativity, Einstein had developed an argument against superluminal velocities that is quite distinct from those featured in the 1905 relativity paper.[6] It is not unlikely that Wien had read Einstein's June 1907 paper and this aroused in him the interest in the question.[7]

In his June 1907 review article, Einstein wrote, "The addition theorem for velocities also yields the interesting conclusion that there cannot exist an effect that can be used for arbitrary signaling and that is propagated faster than light in vacuum."[8]

---

[3] Einstein, 1905, p. 906.
[4] "Einstein on Superluminal Signal Velocities", *The Collected Papers of Albert Einstein* (*CPAE*), *Vol. 5: The Swiss Years: Correspondence, 1902–1914*, Klein, Martin J., Kox, A.J., and Schulmann, Robert (eds.), Princeton: Princeton University Press, 1993, pp. 56-57.
[5] "Einstein on Superluminal Signal Velocities", *CPAE*, Vol. 5, p. 57.
[6] Einstein, Albert, "Über das Relativitätsprinzip und die aus demselben gezogenen Folgerungen", *Jahrbuch der Radioaktivität* 4, 1907, pp. 411-462; 5, 1908, pp. 98-99 (Berichtigungen, errata).
[7] "Einstein on Superluminal Signal Velocities", *CPAE*, Vol. 5, p. 57.
[8] Einstein, 1907, p. 423.

Einstein used the relativistic addition theorem for velocities to show that *a signal* propagating superluminally from cause to effect in the rest frames of those two events will propagate from effect to cause in another frame moving relatively to the first.[9]

"If, as we have assumed", wrote Einstein, [the velocity of propagation of a body relative to a system S] $W > c$, one can always choose **v** such that [time] $T < 0$. This result means that we would have to consider as possible a transfer mechanism whereby the achieved effect would precede the cause. Even though this result, in my opinion, does not contain any contradiction from a purely logical point of view, it conflicts with the character of all our experience to such an extent that this seems sufficient to prove the impossibility of the assumption $W > c$".[10]

The fundamental impossibility of superluminal speeds became controversial because it became clear that in dispersive and absorptive media the *phase velocity* of a plane wave and even the *group velocity* of a superposition of waves could exceed the speed of light. Thus Maxwell's theory seemed to have contradicted relativity.[11]

Wien could not accept superluminal velocities and wrote Einstein a letter in which he used the standard expression for the group velocity to support his case:

$U = V - \lambda(dV/d\lambda)$, where $V$ is the phase velocity and $\lambda$ the wavelength.

Einstein was confronted with a new challenge: Was his 1905 new relativistic addition law of velocities applicable to group velocity in dispersive media? He tried to present to Wien an alternative expression for the group velocity in dispersive media, which was based on his relativistic addition law of velocities:

$U = V/[1+ (\lambda/V)(dV/d\lambda)]$,

$\lambda$ wavelength (in vacuum) $V$ velocity of light (in the medium) [Einstein's notation and words],

which he claimed to be valid for absorptive media and so superluminal velocities would be incompatible with Maxwell's theory.[12]

He wrote Wien, "In my opinion, there is a contradiction with the principle of relativity in conjunction with the principle of the constancy of the velocity of light in the vacuum if for a spec. [ific] metal and specific color [the phase velocity] $U > L$ (velocity of light in vacuum)" And "The propagation of an electromagnetic signal with superluminal velocity is also incompatible with Maxwell's theory of electricity &

---

[9] "Einstein on Superluminal Signal Velocities", *CPAE*, Vol. 5, p. 57.
[10] Einstein, 1907, p. 424.
[11] "Einstein on Superluminal Signal Velocities", *CPAE*, Vol. 5, p. 57.
[12] "Einstein on Superluminal Signal Velocities", *CPAE*, Vol. 5, p. 58.

light. This follows from the results of a study by [Emil] Wiechert that was published in [1900 in the] Lorentz Festschrift. In this study it is shown that one obtains something equivalent to Maxwell's equations if one introduces certain actions-at-a-distance that propagate with the velocity of light L in the vacuum and act from one electric mass to the other".[13]

And the expression that he proposed according to Wiechart's theory contradicted Maxwell's theory as well, "As I have now recognized the letter I sent you yesterday is in need of an amendment and a correction. We must stick with the relation […]" and Einstein came back to the standard expression of the group velocity of the waves, which Wien had proposed.[14]

But Einstein was finally confused, "Unfortunately, almost everything that I reported to you, all too rashly, in my previous letters proved to be false upon closer examination. In fact, the only thing that is to be retained as correct is that Maxwell's theory rules out the possibility of the propagation of a signal with superluminal velocity. For the case of non absorbing bodies or non-absorbing regions of a body, the expression for the group velocity is also, in my opinion, correct. But it is in general difficult to retain the concept of 'group velocity' for absorptive bodies if the absorption coefficient depends on the frequency. My reasoning was faulty because I implicitly assumed that the propagation law of an amplitude must be expressed by a *first order* differential equation."

Einstein was confused because he did not have the correct expression for the group velocity in dispersive media. However, he knew that the only correct thing is that, "according to Wiechart's results it is beyond doubt that our electromagnetic theories of dispersion can never yield superluminal velocity for the propagation of an optical signal."[15] He vacillated between the classical expression that Wien wrote him and his own expression, which was based on his relativistic addition law of velocities.

Finally, Einstein gave up the whole project and he now explained to Wien that extremely complicated investigations would be needed in order to calculate the propagation velocities of optical signals in metals; small wonder that in 1907, Wien was not convinced by Einstein's arguments. He remained concerned with the problem of superluminal velocities.[16]

I wish to thank Prof. John Stachel from the Center for Einstein Studies in Boston University for sitting with me for many hours discussing special and general relativity and their history.

Israel, March, 2012.

---

[13] Einstein to Wien, 23 August or July, 1907, *CPAE*, Vol. 5, Doc. 49.
[14] Einstein to Wien, 25 August or July, 1907, *CPAE*, Vol. 5, Doc. 50.
[15] Einstein to Wien, 29 July, 1907, *CPAE*, Vol. 5, Doc. 51.
[16] "Einstein on Superluminal Signal Velocities", CPAE, Vol. 5, pp. 58-59; Wien to Einstein to Wien, 26 August, 1907, *CPAE*, Vol. 5, Doc. 55.